\input colordvi
\catcode`@=11                                   
\catcode`\|=12                                  
\catcode`\&=4                                   

\newcount\ncols         \ncols=\z@              
\newcount\nrows         \nrows=\z@              
\newcount\curcol        \curcol=\z@             
     
\newdimen\thinsize      \thinsize=0.6pt         
\newdimen\thicksize     \thicksize=1.5pt        

\newif\iftableinfo      \tableinfotrue          
\newif\ifcentertables   \centertablestrue       
%
%
     
\let\plaincr=\cr                        
\let\plainspan=\span                    
\let\plaintab=&                         
\let\lparen=(                           
\let\NX=\noexpand                       

     
\def\ruledtable{\relax                          
    \@BeginRuledTable                           
    \@RuledTable}


\def\@BeginRuledTable{
   \ncols=0\nrows=0                             
   \begingroup                                  
    \offinterlineskip                           
    \def~{\phantom{0}}
    \def\span{\plainspan\omit\relax\colcount\plainspan}
    \let\cr=\crrule                             
    \let\CR=\crthick                            
    \let\nr=\crnorule                           
    \let\|=\Vb                                  
%
%
    \ifx\tablestrut\undefined\relax             
    \else\let\tstrut=\tablestrut\fi             
    \catcode`\|=13 \catcode`\&=13\relax         
    \TableActive                                
    \curcol=1                                   
%
%
    \ifdim\tablewidth>-\maxdimen\relax          %
      \edef\@Halign{\NX\halign to \NX\tablewidth\NX\bgroup\TablePreamble}%
      \tabskip=0pt plus 1fil                    
    \else                                       %
      \edef\@Halign{\NX\halign\NX\bgroup\TablePreamble}%
      \tabskip=0pt                              
    \fi                                         %
%
%
    \ifcentertables                             
       \ifhmode\vskip 0pt\fi                    
       \line\bgroup\hss                         
    \else\hbox\bgroup                           
    \fi}


\long\def\@RuledTable#1\endruledtable{
   \vrule width\thicksize                       
     \vbox{\@Halign                             
       \thickrule                               
       #1\relax                                 
       \tstrut                                  
       \plaincr\thickrule                       
     \egroup}
   \vrule width\thicksize                       
   \ifcentertables\hss\fi\egroup                
  \endgroup                                     
  \global\tablewidth=-\maxdimen                 
  \iftableinfo                                  
      \immediate\write16{[Nrows=\the\nrows, Ncols=\the\ncols]}%
   \fi}
     

\def\TablePreamble{
   \linecount                           
   \TableItem{####}
   \plaintab\plaintab                   
   \TableItem{####}
   \plaincr}


\def\@TableItem#1{
   \hfil\tablespace                             
   #1\relax                                     
   \tablespace\hfil                             
    }%

\def\@tableright#1{
   \hfil\tablespace\relax               
   #1\relax                             
   \tablespace\relax}

\def\@tableleft#1{
   \tablespace\relax                    
   #1\relax                             
   \tablespace\hfil}

\let\TableItem=\@TableItem              
     
\def\RightJustifyTables{\let\TableItem=\@tableright}
\def\LeftJustifyTables{\let\TableItem=\@tableleft}
\def\NoJustifyTables{\let\TableItem=\@TableItem}

\def\LooseTables{\let\tablespace=\quad}
\def\TightTables{\let\tablespace=\space}
\LooseTables                                    

%

\newdimen\tablewidth    \tablewidth=-\maxdimen  


\def\setRuledStrut{
   \dimen@=\baselineskip                        
   \advance\dimen@ by-\normalbaselineskip       
   \ifdim\dimen@<.5ex \dimen@=.5ex\fi           
   \setbox0=\hbox{\lparen}
   \dimen1=\dimen@ \advance\dimen1 by \ht0      
   \dimen2=\dimen@ \advance\dimen2 by \dp0      
   \def\tstrut{\vrule height\dimen1 depth\dimen2 width\z@}%
   }%

\def\tstrut{\vrule height 3.1ex depth 1.2ex width 0pt}


\def\bigitem#1{
   \setbox0=\hbox{#1}
   \dimen1 =\ht0 \dimen2 =\dp0                  
   \dimen@ =\baselines@ve                       
   \advance\dimen@ by-\normalbaselineskip       
   \ifdim\dimen@<.25ex \dimen@=.25ex\fi         
   \advance\dimen1 by \dimen@                   
   \advance\dimen2 by \dimen@                   
   \vrule height\dimen1 depth\dimen2 width\z@   
   \copy0}

     
%

     
\def\nextcolumn#1{
   \plaintab\omit#1\relax\colcount              
   \plaintab}
     
\def\tab{
   \nextcolumn{\relax}}


\def\vb{
   \nextcolumn{\vrule width\thinsize}}

\def\Vb{
   \nextcolumn{\vrule width\thicksize}}


     
{\catcode`\|=13 \let|0
 \catcode`\&=13 \let&0
 \gdef\TableActive{\let|=\vb \let&=\tab}%
}


\def\crrule{\relax                      
   \tstrut                              
   \plaincr\tablerule                   
  }%

\def\crthick{\relax                     
   \tstrut                              
   \plaincr\thickrule                   
  }%
     
\def\crnorule{\relax                    
   \tstrut                              
   \plaincr                             
   }%
   

     
\def\tablerule{\noalign{\hrule height\thinsize depth 0pt}}%
\def\thickrule{\noalign{\hrule height\thicksize depth 0pt}}%


%
%
%
     

\def\linecount{\relax\global\ncols=\curcol      
   \global\curcol=1                             
   \global\advance\nrows by 1\relax}
     
\def\colcount{\relax                            %
   \global\advance\curcol by 1\relax}


\newdimen\parasize      \parasize=4in           

%

%

\def\begintable{\relax                          
    \@BeginRuledTable                           
    \@begintable}

\long\def\@begintable#1\endtable{
   \@RuledTable#1\endruledtable}


\catcode`@=12                                   


\input epsf
\input amssym

\newfam\scrfam
\batchmode\font\tenscr=rsfs10 \errorstopmode
\ifx\tenscr\nullfont
        \message{rsfs script font not available. Replacing with calligraphic.}
        \def\scr{\cal}
\else   
        \font\sevenscr=rsfs7
        \font\fivescr=rsfs5
        \skewchar\tenscr='177 \skewchar\sevenscr='177 \skewchar\fivescr='177
        \textfont\scrfam=\tenscr \scriptfont\scrfam=\sevenscr
        \scriptscriptfont\scrfam=\fivescr
        \def\scr{\fam\scrfam}
        \def\cal{\scr}
\fi
\catcode`\@=11
\newfam\frakfam
\batchmode\font\tenfrak=eufm10 \errorstopmode
\ifx\tenfrak\nullfont
        \message{eufm font not available. Replacing with italic.}
        \def\frak{\it}
\else
	
	\font\sevenfrak=eufm7 \font\fivefrak=eufm5
        
	\textfont\frakfam=\tenfrak
	\scriptfont\frakfam=\sevenfrak \scriptscriptfont\frakfam=\fivefrak
	\def\frak{\fam\frakfam}
\fi
\catcode`\@=\active
\newfam\msbfam
\batchmode\font\twelvemsb=msbm10 scaled\magstep1 \errorstopmode
\ifx\twelvemsb\nullfont\def\Bbb{\bf}
        
	\font\eightbbb=cmb10 at 8pt
	\message{Blackboard bold not available. Replacing with boldface.}
\else   \catcode`\@=11
        
        \font\tenmsb=msbm10 \font\sevenmsb=msbm7 \font\fivemsb=msbm5
        \textfont\msbfam=\tenmsb
        \scriptfont\msbfam=\sevenmsb \scriptscriptfont\msbfam=\fivemsb
        \def\Bbb{\relax\expandafter\Bbb@}
        \def\Bbb@#1{{\Bbb@@{#1}}}
        \def\Bbb@@#1{\fam\msbfam\relax#1}
        \catcode`\@=\active

	\font\eightbbb=msbm8
\fi
        \font\fivemi=cmmi5
        \font\sixmi=cmmi6
        \font\eightrm=cmr8              \def\xrm{\eightrm}
        \font\eightbf=cmbx8             \def\xbf{\eightbf}
        \font\eightit=cmti10 at 8pt     \def\xit{\eightit}

        \font\eighttt=cmtt8

        \font\eightcp=cmcsc8    
                      \def\xold{\eighti}
        \font\eightmi=cmmi8
                     \def\xbold{\eightib}
        \font\teni=cmmi10               \def\old{\teni}
        \font\tencp=cmcsc10

        \font\twelvecp=cmcsc10 scaled\magstep1
        
        \font\fourteenbf=cmbx14
        \font\sixrm=cmr6
        \font\fiverm=cmr5

        \font\eightsy=cmsy8
        \font\sixsy=cmsy6
        \font\eightsl=cmsl8

        \font\sixbf=cmbx6

	 at10pt	
	 at12pt
	 at14pt
	 at16pt
	 at16pt

        \font\twelvebf=cmbx12



\def\xbold{\xbf}
\def\xold{\xrm}


\def\noblackbox{\overfullrule=0pt}
\noblackbox


\def\eightpoint{
\def\rm{\fam0\eightrm}
\textfont0=\eightrm \scriptfont0=\sixrm \scriptscriptfont0=\fiverm
\textfont1=\eightmi  \scriptfont1=\sixmi  \scriptscriptfont1=\fivemi
\textfont2=\eightsy \scriptfont2=\sixsy \scriptscriptfont2=\fivesy
\textfont3=\tenex   \scriptfont3=\tenex \scriptscriptfont3=\tenex
\textfont\itfam=\eightit \def\it{\fam\itfam\eightit}
\textfont\slfam=\eightsl \def\sl{\fam\slfam\eightsl}
\textfont\ttfam=\eighttt \def\tt{\fam\ttfam\eighttt}
\textfont\bffam=\eightbf \scriptfont\bffam=\sixbf 
                         \scriptscriptfont\bffam=\fivebf
                         \def\bf{\fam\bffam\eightbf}
\normalbaselineskip=10pt}



\newtoks\headtext
\headline={\ifnum\pageno=1\hfill\else
	\ifodd\pageno
        \noindent{\eightcp\the\headtext}{ }\dotfill{ }{\old\folio}
	\else\noindent{\old\folio}{ }\dotfill{ }{\eightcp\the\headtext}\fi
	\fi}
\def\makeheadline{\vbox to 0pt{\vss\noindent\the\headline\break
\hbox to\hsize{\hfill}}
        \vskip2\baselineskip}
\newcount\infootnote
\infootnote=0
\newcount\footnotecount
\footnotecount=1
\def\foot#1{\infootnote=1
\footnote{${}^{\the\footnotecount}$}{\vtop{\baselineskip=.75\baselineskip
\advance\hsize by
-\parindent{\eightpoint\rm\hskip-\parindent
#1}\hfill\vskip\parskip}}\infootnote=0\global\advance\footnotecount by
1\hskip2pt}
\newcount\refcount
\refcount=1
\newwrite\refwrite
\def\oldsize{\ifnum\infootnote=1\xold\else\old\fi}
\def\ref#1#2{
	\def#1{{{\oldsize\the\refcount}}\ifnum\the\refcount=1\immediate\openout\refwrite=\jobname.refs\fi\immediate\write\refwrite{\item{[{\xold\the\refcount}]} 
	#2\hfill\par\vskip-2pt}\xdef#1{{\noexpand\oldsize\the\refcount}}\global\advance\refcount by 1}
	}
\def\refout{\eightpoint\catcode`\@=11
        \xrm\immediate\closeout\refwrite
        \vskip2\baselineskip
        {\noindent\twelvecp References}\hfill\vskip\baselineskip
        \baselineskip=.75\baselineskip
        \input\jobname.refs
        \baselineskip=4\baselineskip \divide\baselineskip by 3
        \catcode`\@=\active\rm}

\def\skipref#1{\hbox to15pt{\phantom{#1}\hfill}\hskip-15pt}

\def\hepth#1{\href{http://xxx.lanl.gov/abs/hep-th/#1}{arXiv:\allowbreak
hep-th\slash{\xold#1}}}

\def\arxiv#1#2{\href{http://arxiv.org/abs/#1.#2}{arXiv:\allowbreak
{\xold#1}.{\xold#2}}} 
 
\def\jhep#1#2#3#4{\href{http://jhep.sissa.it/stdsearch?paper=#2\%28#3\%29#4}{J. High Energy Phys. {\xbold #1#2} ({\xold#3}) {\xold#4}}}

\def\FP#1#2#3{Fortsch. Phys. {\xbold#1} ({\xold#2}) {\xold#3}}

\def\JMP#1#2#3{J. Math. Phys. {\xbold#1} ({\xold#2}) {\xold#3}}
\def\JPA#1#2#3{J. Phys. {\xbf A}{\xbold#1} ({\xold#2}) {\xold#3}}

\def\NPB#1#2#3{Nucl. Phys. {\xbf B}{\xbold#1} ({\xold#2}) {\xold#3}}

\def\PLB#1#2#3{Phys. Lett. {\xbf B}{\xbold#1} ({\xold#2}) {\xold#3}}

\def\PRD#1#2#3{Phys. Rev. {\xbf D}{\xbold#1} ({\xold#2}) {\xold#3}}
\def\PRL#1#2#3{Phys. Rev. Lett. {\xbold#1} ({\xold#2}) {\xold#3}}

\newcount\sectioncount
\sectioncount=0
\def\section#1#2{\global\eqcount=0
	\global\subsectioncount=0
        \global\advance\sectioncount by 1
	\ifnum\sectioncount>1
	        \vskip2\baselineskip
	\fi
\noindent{\twelvecp\the\sectioncount. #2}\par\nobreak
       \vskip.5\baselineskip\noindent
        \xdef#1{{\old\the\sectioncount}}}
\newcount\subsectioncount
\def\subsection#1#2{\global\advance\subsectioncount by 1
\vskip.75\baselineskip\noindent\line{\tencp\the\sectioncount.\the\subsectioncount. #2\hfill}\nobreak\vskip.4\baselineskip\nobreak\noindent\xdef#1{{\old\the\sectioncount}.{\old\the\subsectioncount}}}
\def\immediatesubsection#1#2{\global\advance\subsectioncount by 1
\vskip-\baselineskip\noindent
\line{\tencp\the\sectioncount.\the\subsectioncount. #2\hfill}
	\vskip.5\baselineskip\noindent
	\xdef#1{{\old\the\sectioncount}.{\old\the\subsectioncount}}}
\newcount\subsubsectioncount
\def\subsubsection#1#2{\global\advance\subsubsectioncount by 1
\vskip.75\baselineskip\noindent\line{\tencp\the\sectioncount.\the\subsectioncount.\the\subsubsectioncount. #2\hfill}\nobreak\vskip.4\baselineskip\nobreak\noindent\xdef#1{{\old\the\sectioncount}.{\old\the\subsectioncount}.{\old\the\subsubsectioncount}}}
\newcount\appendixcount
\appendixcount=0
\def\appendix#1{\global\eqcount=0
        \global\advance\appendixcount by 1
        \vskip2\baselineskip\noindent
        \ifnum\the\appendixcount=1
        {\twelvecp Appendix A: #1}\par\nobreak
                        \vskip.5\baselineskip\noindent\fi
        \ifnum\the\appendixcount=2
        {\twelvecp Appendix B: #1}\par\nobreak
                        \vskip.5\baselineskip\noindent\fi
        \ifnum\the\appendixcount=3
        {\twelvecp Appendix C: #1}\par\nobreak
                        \vskip.5\baselineskip\noindent\fi}
\def\acknowledgements{\immediate\write\contentswrite{\item{}\hbox
        to\contentlength{Acknowledgements\dotfill\the\pageno}}
        \vskip2\baselineskip\noindent
        \underbar{\it Acknowledgements:}\ }
\newcount\eqcount
\eqcount=0
\def\Eqn#1{\global\advance\eqcount by 1
\ifnum\the\sectioncount=0
	\xdef#1{{\noexpand\oldsize\the\eqcount}}
	\eqno({\oldstyle\the\eqcount})
\else
        \ifnum\the\appendixcount=0
\xdef#1{{\noexpand\oldsize\the\sectioncount}.{\noexpand\oldsize\the\eqcount}}
                \eqno({\oldstyle\the\sectioncount}.{\oldstyle\the\eqcount})\fi
        \ifnum\the\appendixcount=1
	        \xdef#1{{\noexpand\oldstyle A}.{\noexpand\oldstyle\the\eqcount}}
                \eqno({\oldstyle A}.{\oldstyle\the\eqcount})\fi
        \ifnum\the\appendixcount=2
	        \xdef#1{{\noexpand\oldstyle B}.{\noexpand\oldstyle\the\eqcount}}
                \eqno({\oldstyle B}.{\oldstyle\the\eqcount})\fi
        \ifnum\the\appendixcount=3
	        \xdef#1{{\noexpand\oldstyle C}.{\noexpand\oldstyle\the\eqcount}}
                \eqno({\oldstyle C}.{\oldstyle\the\eqcount})\fi
\fi}
\def\eqn{\global\advance\eqcount by 1
\ifnum\the\sectioncount=0
	\eqno({\oldstyle\the\eqcount})
\else
        \ifnum\the\appendixcount=0
                \eqno({\oldstyle\the\sectioncount}.{\oldstyle\the\eqcount})\fi
        \ifnum\the\appendixcount=1
                \eqno({\oldstyle A}.{\oldstyle\the\eqcount})\fi
        \ifnum\the\appendixcount=2
                \eqno({\oldstyle B}.{\oldstyle\the\eqcount})\fi
        \ifnum\the\appendixcount=3
                \eqno({\oldstyle C}.{\oldstyle\the\eqcount})\fi
\fi}
\def\multi{\global\advance\eqcount by 1}
\def\multieqn#1{({\oldstyle\the\sectioncount}.{\oldstyle\the\eqcount}\hbox{#1})}
\def\multiEqn#1#2{\xdef#1{{\oldstyle\the\sectioncount}.{\old\the\eqcount}#2}
        ({\oldstyle\the\sectioncount}.{\oldstyle\the\eqcount}\hbox{#2})}
\def\multiEqnAll#1{\xdef#1{{\oldstyle\the\sectioncount}.{\old\the\eqcount}}}
\newcount\tablecount
\tablecount=0
\def\Table#1#2#3{\global\advance\tablecount by 1
\immediate\write\intrefwrite{\def\noexpand#1{{\noexpand\oldsize\the\tablecount}}}
       \vtop{\vskip2\parskip
       \centerline{#2}
       \vskip5\parskip
       \centerline{\it Table \the\tablecount: #3}
       \vskip2\parskip}}
\newcount\figurecount
\figurecount=0
\def\Figure#1#2#3{\global\advance\figurecount by 1
\immediate\write\intrefwrite{\def\noexpand#1{{\noexpand\oldsize\the\figurecount}}}
       \vtop{\vskip2\parskip
       \centerline{#2}
       \vskip4\parskip
       \centerline{\it Figure \the\figurecount: #3}
       \vskip3\parskip}}
\def\TextFigure#1#2#3{\global\advance\figurecount by 1
\immediate\write\intrefwrite{\def\noexpand#1{{\noexpand\oldsize\the\figurecount}}}
       \vtop{\vskip2\parskip
       \centerline{#2}
       \vskip4\parskip
       {\narrower\noindent\it Figure \the\figurecount: #3\smallskip}
       \vskip3\parskip}}
\newtoks\url
\def\Href#1#2{\catcode`\#=12\url={#1}\catcode`\#=\active#2}
\def\href#1#2{{#2}}

\parskip=3.5pt plus .3pt minus .3pt
\baselineskip=14pt plus .1pt minus .05pt
\lineskip=.5pt plus .05pt minus .05pt
\lineskiplimit=.5pt
\abovedisplayskip=18pt plus 4pt minus 2pt
\belowdisplayskip=\abovedisplayskip
\hsize=14cm
\vsize=19cm
\hoffset=1.5cm
\voffset=1.8cm
\frenchspacing
\footline={}
\raggedbottom

\newskip\origparindent
\origparindent=\parindent

\def\*{\partial}

\def\komma{\;\,,}

\def\={\!=\!}
\def\small#1{{\hbox{$#1$}}}

\def\fraction#1{\small{1\over#1}}
\def\fr{\fraction}
\def\Fraction#1#2{\small{#1\over#2}}
\def\Fr{\Fraction}

\def\eg{{\it e.g.}}

\def\a{\alpha}
\def\b{\beta}

\def\RR{{\Bbb R}}


\def\appendix#1#2{\global\eqcount=0
        \global\advance\appendixcount by 1
        \vskip2\baselineskip\noindent
        \ifnum\the\appendixcount=1
        \immediate\write\intrefwrite{\def\noexpand#1{A}}
        {\twelvecp Appendix A: #2}\par\nobreak
                        \vskip.5\baselineskip\noindent\fi
        \ifnum\the\appendixcount=2
        {\twelvecp Appendix B: #2}\par\nobreak
                        \vskip.5\baselineskip\noindent\fi
        \ifnum\the\appendixcount=3
        {\twelvecp Appendix C: #2}\par\nobreak
                        \vskip.5\baselineskip\noindent\fi}

\def\textfrac#1#2{\raise .45ex\hbox{\the\scriptfont0 #1}\nobreak\hskip-1pt/\hskip-1pt\hbox{\the\scriptfont0 #2}}

\def\LL{{\cal L}}


\def\frac{\Fr}

\def\mathbb{\Bbb}

\def\ZZ{{\Bbb Z}}



\def\LL{{\cal L}}

\def\LL{{\cal L}}

\def\LL{{\cal L}}




\catcode`@=11
\def\openupnormal{\afterassignment\@penupnormal\dimen@=}
\def\@penupnormal{\advance\normallineskip\dimen@
  \advance\normalbaselineskip\dimen@
  \advance\normallineskiplimit\dimen@}
\catcode`@=12

\def\EqMatrix{\let\quad\enspace\openupnormal6pt\matrix}



\def\textfrac#1#2{\raise .45ex\hbox{\the\scriptfont0 #1}\nobreak\hskip-1pt/\hskip-1pt\hbox{\the\scriptfont0 #2}}

\def\LL{{\cal L}}


\def\frac{\Fr}

\def\mathbb{\Bbb}

\newskip\scrskip
\scrskip=-.6pt plus 0pt minus .1pt


\newwrite\intrefwrite
\immediate\openout\intrefwrite=\jobname.intref

\newwrite\contentswrite

\newdimen\sublength
\sublength=\hsize 
\advance\sublength by -\parindent

\newdimen\contentlength
\contentlength=\sublength

\advance\sublength by -\parindent

\def\section#1#2{\global\eqcount=0
	\global\subsectioncount=0
        \global\advance\sectioncount by 1
\ifnum\the\sectioncount=1\immediate\openout\contentswrite=\jobname.contents\fi
\immediate\write\contentswrite{\item{\the\sectioncount.}\hbox to\contentlength{#2\dotfill\the\pageno}}
	\ifnum\sectioncount>1
	        \vskip2\baselineskip
	\fi
\immediate\write\intrefwrite{\def\noexpand#1{{\noexpand\oldsize\the\sectioncount}}}\noindent{\twelvecp\the\sectioncount. #2}\par\nobreak
       \vskip.5\baselineskip\noindent}

\def\subsection#1#2{\global\advance\subsectioncount by 1
\immediate\write\contentswrite{\itemitem{\the\sectioncount.\the\subsectioncount.}\hbox
to\sublength{#2\dotfill\the\pageno}}
\immediate\write\intrefwrite{\def\noexpand#1{{\noexpand\oldsize\the\sectioncount}.{\noexpand\oldsize\the\subsectioncount}}}\vskip.75\baselineskip\noindent\line{\tencp\the\sectioncount.\the\subsectioncount. #2\hfill}\nobreak\vskip.4\baselineskip\nobreak\noindent}

\def\immediatesubsection#1#2{\global\advance\subsectioncount by 1
\immediate\write\contentswrite{\itemitem{\the\sectioncount.\the\subsectioncount.}\hbox
to\sublength{#2\dotfill\the\pageno}}
\immediate\write\intrefwrite{\def\noexpand#1{{\noexpand\oldsize\the\sectioncount}.{\noexpand\oldsize\the\subsectioncount}}}
\vskip-\baselineskip\noindent
\line{\tencp\the\sectioncount.\the\subsectioncount. #2\hfill}
	\vskip.5\baselineskip\noindent}

\def\contentsout{\catcode`\@=11
        \vskip2\baselineskip
        {\noindent\twelvecp Contents}\hfill\vskip\baselineskip
        \input\jobname.contents
        \catcode`\@=\active\rm
\vskip3\baselineskip
}

\def\refout{\eightpoint\catcode`\@=11
        \immediate\write\contentswrite{\item{}\hbox to\contentlength{References\dotfill\the\pageno}}
        \xrm\immediate\closeout\refwrite
        \vskip2\baselineskip
        {\noindent\twelvecp References}\hfill\vskip\baselineskip
        \baselineskip=.75\baselineskip
        \input\jobname.refs
        \baselineskip=4\baselineskip \divide\baselineskip by 3
        \catcode`\@=\active\rm}


\ref\Tseytlin{A.A.~Tseytlin,
  {\xit ``Duality symmetric closed string theory and interacting
  chiral scalars''}, 
  \NPB{350}{1991}{395}.}

\ref\SiegelI{W.~Siegel,
  {\xit ``Two vierbein formalism for string inspired axionic gravity''},
  \PRD{47}{1993}{5453}
  [\hepth{9302036}].}

\ref\SiegelII{ W.~Siegel,
  {\xit ``Superspace duality in low-energy superstrings''},
  \PRD{48}{1993}{2826}
  [\hepth{9305073}].}

\ref\SiegelIII{W.~Siegel,
  {\xit ``Manifest duality in low-energy superstrings''},
  in Berkeley 1993, Proceedings, Strings '93 353
  [\hepth{9308133}].}

\ref\HullDoubled{C.M. Hull, {\xit ``Doubled geometry and
T-folds''}, \jhep{07}{07}{2007}{080}
[\hepth{0605149}].}

\ref\HullT{C.M. Hull, {\xit ``A geometry for non-geometric string
backgrounds''}, \jhep{05}{10}{2005}{065} [\hepth{0406102}].}

\ref\HullM{C.M. Hull, {\xit ``Generalised geometry for M-theory''},
\jhep{07}{07}{2007}{079} [\hepth{0701203}].}

\ref\HullZwiebachDFT{C. Hull and B. Zwiebach, {\xit ``Double field
theory''}, \jhep{09}{09}{2009}{99} [\arxiv{0904}{4664}].}

\ref\HohmHullZwiebachI{O. Hohm, C.M. Hull and B. Zwiebach, {\xit ``Background
independent action for double field
theory''}, \jhep{10}{07}{2010}{016} [\arxiv{1003}{5027}].}

\ref\HohmHullZwiebachII{O. Hohm, C.M. Hull and B. Zwiebach, {\xit
``Generalized metric formulation of double field theory''},
\jhep{10}{08}{2010}{008} [\arxiv{1006}{4823}].} 

\ref\HohmKwak{O. Hohm and S.K. Kwak, {\xit ``$N=1$ supersymmetric
double field theory''}, \jhep{12}{03}{2012}{080} [\arxiv{1111}{7293}].}

\ref\HohmKwakFrame{O. Hohm and S.K. Kwak, {\xit ``Frame-like geometry
of double field theory''}, \JPA{44}{2011}{085404} [\arxiv{1011}{4101}].}

\ref\JeonLeeParkI{I. Jeon, K. Lee and J.-H. Park, {\xit ``Differential
geometry with a projection: Application to double field theory''},
\jhep{11}{04}{2011}{014} [\arxiv{1011}{1324}].}

\ref\JeonLeeParkII{I. Jeon, K. Lee and J.-H. Park, {\xit ``Stringy
differential geometry, beyond Riemann''}, 
\PRD{84}{2011}{044022} [\arxiv{1105}{6294}].}

\ref\JeonLeeParkIII{I. Jeon, K. Lee and J.-H. Park, {\xit
``Supersymmetric double field theory: stringy reformulation of supergravity''},
\PRD{85}{2012}{081501} [\arxiv{1112}{0069}].}

\ref\HohmZwiebachLarge{O. Hohm and B. Zwiebach, {\xit ``Large gauge
transformations in double field theory''}, \jhep{13}{02}{2013}{075}
[\arxiv{1207}{4198}].} 

\ref\Park{J.-H.~Park,
  {\xit ``Comments on double field theory and diffeomorphisms''},
  \jhep{13}{06}{2013}{098}
  [\arxiv{1304}{5946}].}

\ref\BermanCederwallPerry{D.S. Berman, M. Cederwall and M.J. Perry,
{\xit ``Global aspects of double geometry''}, 
\jhep{14}{09}{2014}{66} [\arxiv{1401}{1311}].}

\ref\PachecoWaldram{P.P. Pacheco and D. Waldram, {\xit ``M-theory,
exceptional generalised geometry and superpotentials''},
\jhep{08}{09}{2008}{123} [\arxiv{0804}{1362}].}

\ref\Hillmann{C. Hillmann, {\xit ``Generalized $E_{7(7)}$ coset
dynamics and $D=11$ supergravity''}, \jhep{09}{03}{2009}{135}
[\arxiv{0901}{1581}].}

\ref\BermanPerryGen{D.S. Berman and M.J. Perry, {\xit ``Generalised
geometry and M-theory''}, \jhep{11}{06}{2011}{074} [\arxiv{1008}{1763}].}    

\ref\BermanGodazgarPerry{D.S. Berman, H. Godazgar and M.J. Perry,
{\xit ``SO(5,5) duality in M-theory and generalized geometry''},
\PLB{700}{2011}{65} [\arxiv{1103}{5733}].} 

\ref\BermanMusaevPerry{D.S. Berman, E.T. Musaev and M.J. Perry,
{\xit ``Boundary terms in generalized geometry and doubled field theory''},
\PLB{706}{2011}{228} [\arxiv{1110}{97}].} 

\ref\BermanGodazgarGodazgarPerry{D.S. Berman, H. Godazgar, M. Godazgar  
and M.J. Perry,
{\xit ``The local symmetries of M-theory and their formulation in
generalised geometry''}, \jhep{12}{01}{2012}{012}
[\arxiv{1110}{3930}].} 

\ref\BermanGodazgarPerryWest{D.S. Berman, H. Godazgar, M.J. Perry and
P. West,
{\xit ``Duality invariant actions and generalised geometry''}, 
\jhep{12}{02}{2012}{108} [\arxiv{1111}{0459}].} 

\ref\CoimbraStricklandWaldram{A. Coimbra, C. Strickland-Constable and
D. Waldram, {\xit ``$E_{d(d)}\times\hbox{\eightbbb R}^+$ generalised geometry,
connections and M theory'' }, \jhep{14}{02}{2014}{054} [\arxiv{1112}{3989}].} 

\ref\CoimbraStricklandWaldramII{A. Coimbra, C. Strickland-Constable and
D. Waldram, {\xit ``Supergravity as generalised geometry II:
$E_{d(d)}\times\hbox{\eightbbb R}^+$ and M theory''}, 
\jhep{14}{03}{2014}{019} [\arxiv{1212}{1586}].}  

\ref\JeonLeeParkSuh{I. Jeon, K. Lee, J.-H. Park and Y. Suh, {\xit
``Stringy unification of Type IIA and IIB supergravities under N=2
D=10 supersymmetric double field theory''}, \PLB{723}{2013}{245}
[\arxiv{1210}{5048}].} 

\ref\JeonLeeParkRR{I. Jeon, K. Lee and J.-H. Park, {\xit
``Ramond--Ramond cohomology and O(D,D) T-duality''},
\jhep{12}{09}{2012}{079} [\arxiv{1206}{3478}].} 

\ref\BermanCederwallKleinschmidtThompson{D.S. Berman, M. Cederwall,
A. Kleinschmidt and D.C. Thompson, {\xit ``The gauge structure of
generalised diffeomorphisms''}, \jhep{13}{01}{2013}{64} [\arxiv{1208}{5884}].}

\ref\ParkSuh{J.-H. Park and Y. Suh, {\xit ``U-geometry: SL(5)''},
\jhep{14}{06}{2014}{102} [\arxiv{1302}{1652}].} 

\ref\CederwallI{M.~Cederwall, J.~Edlund and A.~Karlsson,
  {\xit ``Exceptional geometry and tensor fields''},
  \jhep{13}{07}{2013}{028}
  [\arxiv{1302}{6736}].}

\ref\CederwallII{ M.~Cederwall,
  {\xit ``Non-gravitational exceptional supermultiplets''},
  \jhep{13}{07}{2013}{025}
  [\arxiv{1302}{6737}].}

\ref\HohmSamtlebenI{O.~Hohm and H.~Samtleben,
  {\xit ``Exceptional field theory I: $E_{6(6)}$ covariant form of
  M-theory and type IIB''}, 
  \PRD{89}{2014}{066016} [\arxiv{1312}{0614}].}

\ref\HohmSamtlebenII{O.~Hohm and H.~Samtleben,
  {\xit ``Exceptional field theory II: $E_{7(7)}$''},
  \PRD{89}{2014}{066016} [\arxiv{1312}{4542}].}

\ref\HohmSamtlebenIII{O. Hohm and H. Samtleben, {\xit ``Exceptional field
theory III: $E_{8(8)}$''}, \PRD{90}{2014}{066002} [\arxiv{1406}{3348}].}

\ref\KachruNew{S. Kachru, M.B. Schulz, P.K. Tripathy and S.P. Trivedi,
{\xit ``New supersymmetric string compactifications''}, 
\jhep{03}{03}{2003}{061} [\hepth{0211182}].}

\ref\Condeescu{C. Condeescu, I. Florakis, C. Kounnas and D. L\"ust, 
{\xit ``Gauged supergravities and non-geometric $Q$/$R$-fluxes from
asymmetric orbifold CFT's''}, 
\jhep{13}{10}{2013}{057} [\arxiv{1307}{0999}].}


\ref\HasslerLust{F. Hassler and D. L\"ust, {\xit ``Consistent
compactification of double field theory on non-geometric flux
backgrounds''}, \jhep{14}{05}{2014}{085} [\arxiv{1401}{5068}].}

\ref\CederwallGeometryBehind{M. Cederwall, {\xit ``The geometry behind
double geometry''}, 
\jhep{14}{09}{2014}{70} [\arxiv{1402}{2513}].}

\ref\CederwallDuality{M. Cederwall, {\xit ``T-duality and
non-geometric solutions from double geometry''}, \FP{62}{2014}{942}
[\arxiv{1409}{4463}].} 

\ref\CederwallRosabal{M. Cederwall and J.A. Rosabal, ``$E_8$
geometry'', \jhep{15}{07}{2015}{007}, [\arxiv{1504}{04843}].}

\ref\HohmKwakZwiebachI{O. Hohm, S.K. Kwak and B. Zwiebach, {\xit
``Unification of type II strings and T-duality''}, \PRL{107}{2011}{171603}
[\arxiv{1106}{5452}].}  

\ref\HohmKwakZwiebachII{O. Hohm, S.K. Kwak and B. Zwiebach, {\xit
``Double field theory of type II strings''}, \jhep{11}{09}{2011}{013}
[\arxiv{1107}{0008}].}  

\ref\HohmZwiebachGeometry{O. Hohm and B. Zwiebach, {\xit ``Towards an
invariant geometry of double field theory''}, \arxiv{1212}{1736}.} 

\ref\CartanSpinors{E. Cartan, {\xit ``Le\hskip.5pt,\hskip-3.5pt cons sur
la th\'eorie des spineurs''} (Hermann, Paris, 1937).}

\ref\KacBook{V. G. Kac, {\xit ``Infinite-dimensional Lie
algebras''}, Cambridge Univ. Press
({\xold1990}).}

\ref\KacSuperalgebras{V.G. Kac, {\xit ``Classification of simple Lie
superalgebras''}, Funktsional. Anal. i Prilozhen. {\xbold9}
({\xold1975}) {\xold91}.}

\ref\CederwallExceptionalTwistors{M. Cederwall, {\xit ``Twistors and
supertwistors for exceptional field
theory''}, \jhep{15}{12}{2015}{123} [\arxiv{1510}{02298}].}

\ref\CederwallDoubleSuperGeometry{M. Cederwall, {\xit ``Double
supergeometry''}, \jhep{16}{06}{2016}{155} [\arxiv{1603}{04684}].}

\ref\CederwallPalmkvistBorcherds{M. Cederwall and J. Palmkvist, {\xit
``Superalgebras, constraints and partition functions''},
\jhep{08}{15}{2015}{36} [\arxiv{1503}{06215}].}

\ref\PalmkvistBorcherds{J. Palmkvist, {\xit
``Exceptional geometry and Borcherds superalgebras''},
\jhep{15}{11}{2015}{032} [\arxiv{1507}{08828}].}

\ref\StricklandConstable{C. Strickland-Constable,
  {\xit ``Subsectors, Dynkin diagrams and new generalised geometries''},
  \jhep{17}{08}{2017}{144}
  [\arxiv{1310}{4196}].}

\ref\Baraglia{D. Baraglia. {\xit ``Leibniz algebroids, twistings and
  exceptional generalized geometry''}, J. Geom. 
         Phys. {\xbf62} (2012) 903 [\arxiv{1101}{0856}].}

\ref\ENinePaper{G. Bossard, M. Cederwall, A. Kleinschmidt,
J. Palmkvist and H. Samtleben, {\xit ``Generalised diffeomorphisms for
$E_9$''}, \arxiv{1708}{08936}.}

\ref\HohmSamtlebenIII{O. Hohm and H. Samtleben, {\xit ``Exceptional field
theory III: $E_{8(8)}$''}, \PRD{90}{2014}{066002} [\arxiv{1406}{3348}].}

\ref\CederwallRosabal{M. Cederwall and J.A. Rosabal, ``$E_8$
geometry'', \jhep{15}{07}{2015}{007}, [\arxiv{1504}{04843}].}

\ref\PetersonKac{D.H. Peterson and V.G. Kac, {\xit ``Infinite flag
varieties and conjugacy theorems''}, Proc. Natl. Acad. Sci. {\xbf80}
(1983) 1778.}

\ref\ParkSuhN{J.-H. Park and Y. Suh, {\xit ``U-gravity: SL(N)''},
\jhep{13}{04}{2013}{102} [\arxiv{1402}{5027}].]}

\ref\HohmZwiebachLarge{O. Hohm and B. Zwiebach, {\xit ``Large gauge
transformations in double field theory''}, \arxiv{1207}{4198}.}

\ref\PalmkvistTensor{J. Palmkvist, {\xit ``The tensor hierarchy
algebra''}, \JMP{55}{2014}{011701} [\arxiv{1305}{0018}].}

\ref\CarboneCederwallPalmkvist{L. Carbone, M. Cederwall and
J. Palmkvist, {\xit ``Generators and relations for Lie superalgebras
of Cartan type''}, in preparation.}

\ref\BeyondEEleven{G. Bossard, A. Kleinschmidt, J. Palmkvist,
C.N. Pope and E. Sezgin, {\xit ``Beyond $E_{11}$''},
\jhep{17}{05}{2017}{020} [\arxiv{1703}{01305}].}

\ref\HitchinLectures{N. Hitchin, {``\xit Lectures on generalized
geometry''}, \arxiv{1010}{2526}.}

\ref\deWitNicolaiSamtleben{B.~de Wit, H.~Nicolai and H.~Samtleben,
{\xit ``Gauged supergravities, tensor hierarchies, and M-theory},
\jhep{08}{02}{2008}{044}
[\arxiv{0801}{1294}].}

\ref\KacSuperalgebras{V.~Kac, {\xit ``Lie superalgebras''}, Adv. Math. {\xbf26}
(1977) 8.}

\ref\KacFiniteGrowth{V.~Kac, {\xit ``Simple irreducible graded Lie algebras of finite growth''}, Math. USSR Izv. {\xbf2}
(1968) 1271.}

\ref\RayBook{U.~Ray, {\xit ``Automorphic forms and Lie
superalgebras''}, Springer
({\xold2006}).}

\ref\BossardKleinSchmidtLoops{G. Bossard and A. Kleinschmidt, {\xit
``Loops in exceptional field theory''}, \jhep{16}{01}{2016}{164} [\arxiv{1510}{07859}].}

\ref\HohmSamtlebenEhlers{O. Hohm and H. Samtleben, {\xit ``U-duality
covariant gravity''}, \jhep{13}{09}{2013}{080} [\arxiv{1307}{0509}].}

\ref\HohmMusaevSamtleben{O. Hohm, E.T. Musaev and H. Samtleben, {\xit
``$O(d+1,d+1)$ enhanced double field
theory''}, \jhep{17}{10}{2017}{086} [\arxiv{1707.06693}].}

\ref\BlumenhagenHasslerLust{R. Blumenhagen, F. Hassler and D. L\"ust,
{\xit ``Double field theory on group manifolds''},
\jhep{15}{02}{2015}{001} [\arxiv{1410}{6374}].}

\ref\BlumenhagenBosqueHasslerLust{R. Blumenhagen, P. du Bosque,
F. Hassler and D. L\"ust, 
{\xit ``Generalized metric formulation of double field theory on group
manifolds''}, \arxiv{1502}{02428}.}

\ref\HasslerTopology{F. Hassler, {\xit ``The topology of double field
theory''}, \arxiv{1611}{07978}.}

\ref\HasslerTopology{F. Hassler, {\xit ``Poisson--Lie T-duality in
double field theory''}, \arxiv{1707}{08624}.}

\ref\CederwallPalmkvistExtendedGeometry{M. Cederwall and J. Palmkvist,
{\xit ``Extended geometries''}, \arxiv{1711}{07694}.}

\ref\HohmZwiebachLInfinity{O. Hohm and B. Zwiebach, {\xit ``$L_\infty$
algebras and field theory''}, \FP{65}{2017}{1700014} [\arxiv{1701}{08824}].]}

\ref\BermanCederwallStrickland{D.S. Berman, M. Cederwall and
C. Strickland-Constable, work in progress.}

\ref\DamourHenneauxNicolaiII{T. Damour, M. Henneaux and H. Nicolai,
{\xit ``$E_{10}$ and a 'small tension expansion' of M theory''},
\PRL{89}{2002}{221601} [\hepth{0207267}].}


\def\LL{{\cal L}}
\def\RR{{\Bbb R}}
\def\ZZ{{\Bbb Z}}

\def\d{\delta}






\headline={\hfill}



\centerline{\fourteenbf Algebraic structures in exceptional geometry}

%

\vskip6\parskip

\centerline{\twelvebf Martin Cederwall\foot{martin.cederwall@chalmers.se}}

\vskip4\parskip

\centerline{Dept. of Physics, Chalmers University
of Technology}

\centerline{SE 412 96, 
Gothenburg, Sweden}

\vskip8\parskip






\centerline{\tencp Abstract}
\noindent Exceptional field theory (EFT)
gives a geometric underpinning of the U-duality symmetries of
M-theory. In this talk I give an overview of the surprisingly rich
algebraic structures which naturally appear in the context of
EFT. This includes Borcherds superalgebras, Cartan type
superalgebras (tensor hierarchy algebras) and $L_\infty$ algebras.
This is the written version of a talk based mainly on refs.
[\PalmkvistBorcherds\skipref\PalmkvistTensor\skipref\ENinePaper\skipref\CederwallPalmkvistExtendedGeometry\skipref\CarboneCederwallPalmkvist--\BermanCederwallStrickland],
presented
at ISQS25, Prague, June 2017,
at QTS-10/LT-12, Varna, June 2017, 
at SQS 2017, Dubna, Aug. 2017,
and at M$\cap\Phi$9, Belgrade, Sept. 2017.

\vskip10\parskip


\noindent Duality symmetries in string theory/M-theory mix gravitational and
non-gravitational fields. 
Manifestation of such symmetries calls for a generalisation of the
concept of geometry.
It has been proposed that the compactifying space (torus) is enlarged
to accommodate momenta (representing momenta and brane windings) in
modules of a duality group.  
This leads to {\it double geometry}
[\Tseytlin\skipref\SiegelI\skipref\SiegelIII\skipref\HitchinLectures\skipref\HullT\skipref\HullDoubled\skipref\HullZwiebachDFT\skipref\HohmHullZwiebachI\skipref\HohmHullZwiebachII\skipref\HohmKwakFrame\skipref\HohmKwak\skipref\JeonLeeParkI\skipref\JeonLeeParkII\skipref\JeonLeeParkIII\skipref\HohmZwiebachGeometry\skipref\HohmKwakZwiebachII\skipref\JeonLeeParkSuh\skipref\JeonLeeParkRR\skipref\HohmZwiebachLarge\skipref\Park\skipref\BermanCederwallPerry\skipref\CederwallGeometryBehind\skipref\CederwallDuality\skipref\BlumenhagenHasslerLust--\BlumenhagenBosqueHasslerLust]
in the context of T-duality, 
 and {\it exceptional geometry} [\HullM\skipref\PachecoWaldram\skipref\Hillmann\skipref\BermanPerryGen\skipref\BermanGodazgarPerry\skipref\BermanGodazgarGodazgarPerry\skipref\BermanGodazgarPerryWest\skipref\CoimbraStricklandWaldram\skipref\CoimbraStricklandWaldramII\skipref\BermanCederwallKleinschmidtThompson\skipref\ParkSuh\skipref\CederwallI\skipref\CederwallII\skipref\HohmSamtlebenEhlers\skipref\HohmSamtlebenI\skipref\HohmSamtlebenII\skipref\HohmSamtlebenIII\skipref\CederwallRosabal\skipref\CederwallExceptionalTwistors\skipref\BossardKleinSchmidtLoops--\HohmMusaevSamtleben]
in the context of U-duality. These classes of models are special cases
of {\it extended geometries}, and can be treated in a unified manner
[\CederwallPalmkvistExtendedGeometry].
The duality group is in a certain sense present already in the uncompactified
theory. It becomes ``geometrised''.


In the present talk, I will

\item{$\bullet$}Describe the basics of extended geometry, with focus
on the gauge transformations;



\item{$\bullet$}Describe the appearance of Borcherds superalgebras and
Cartan-type superalgebras (tensor hierarchy superalgebras);

\item{$\bullet$}Indicate why $L_\infty$ algebras provide a good
framework for describing the gauge symmetries.

\item{$\bullet$}{Point out some questions and directions.}


The focus will thus be on algebraic aspects, and less on geometric ones.

\vfill\eject

Consider compactification from $11$ to $11-n$ dimensions on $T^n$. 
As is well known, fields and charges fall into modules of 
$E_{n(n)}$.

\vskip2\parskip
\ruledtable
$n$|$E_{n(n)}$|$R_1$\crthick
$3$|$SL(3)\times SL(2)$|$({\bf3},{\bf2})$\cr
$4$|$SL(5)$|$\bf10$\cr
$5$|$Spin(5,5)$|$\bf16$\cr
$6$|$E_{6(6)}$|$\bf27$\cr
$7$|$E_{7(7)}$|$\bf56$\cr
$8$|$E_{8(8)}$|$\,\,\bf248$\cr
$9$|$E_{9(9)}$|$\,\,\hbox{fund}$
\endruledtable
\vskip2\parskip
\centerline{\it Table 1: A list of U-duality groups.}
\vskip2\parskip

\Figure\ROneFigure{\epsffile{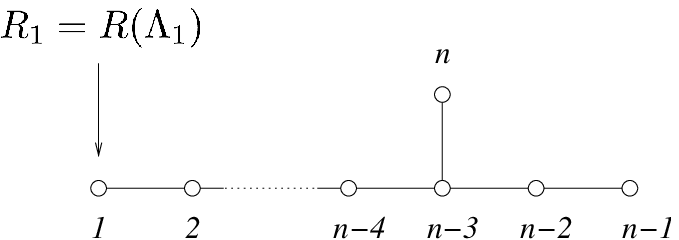}}{The module $R_1$.}

To be explicit, take $n=7$ as an example.
The gauge parameters $\xi^M$ in ${\bf56}$ of $E_7$ decompose as:
$$
\matrix{\xi^m&&\lambda_{mn}&&\tilde\lambda_{mnpqr}
              &&\tilde\xi_{m,n_1\ldots n_7}&\leftarrow&\xi^M\cr
       7&+&21&+&21&+&7&=&56\cr}\eqn
$$
We recognise the parameters for diffeomorphisms, gauge transformations
of the 3-form and dual 6-form and a parameter for ``dual diffeomorphisms''.
The scalar fields are in the coset
$E_{7(7)}/K(E_{7(7)})=E_{7(7)}/(SU(8)/\ZZ_2)$. The dimension of
coset is: $133-63=70$,
and it is parametrised by
$$
\matrix{g_{mn}&&C_{mnp}&&\tilde C_{mnpqrs}&&\leftarrow G_{MN}\cr
       28&+&35&+&7&=&70\cr}\eqn
$$
From the point of view of $N=8$ supergravity in $D=4$, this is the
scalar field coset. Now it becomes a generalised metric.
There are also mixed fields (generalised graviphotons): 1-forms in $R_1$, etc.

The situation for T-duality is simpler.
Compactification from 10 to $10-d$ dimensions gives the (continuous) T-duality 
group $O(d,d)$. The momenta are complemented with string windings to
form the $2d$-dimensional module


Note that the continuous duality group is not to be seen as a global symmetry.
Discrete duality transformations in $O(d,d;\ZZ)$ or $E_{n(n)}(\ZZ)$
arise as symmetries in certain backgrounds, roughly as the mapping
class group $SL(n;\ZZ)$ arises as discrete isometries of a torus.
The r\^ole of the continuous versions of the 
duality groups is analogous to 
that of $GL(n)$ in ordinary geometry (gravity).

One has to decide how tensors transform.
The generic recipe is to mimic the Lie derivative for ordinary
diffeomorphisms: 
$$
L_UV^m=U^n\*_nV^m-\Red{\*_nU^m}V^n\;.\eqn
$$
The first term is a transport term, and the second one a 
${\frak gl}$ transformation, with parameter in red.

In the case of U-duality, the role of $GL$ is assumed by 
$E_{n(n)}\times\RR^+$, and 
$$
\eqalign{
\LL_UV^M&=L_UV^M+Y^{MN}{}_{PQ}\*_NU^PV^Q\cr
&=U^N\*_NV^M+\Red{Z^{MN}{}_{PQ}\*_NU^P}V^Q\;,\cr
}\eqn
$$
where $Z^{MN}{}_{PQ}=-\a_nP_{\hbox{\sixrm
    adj}}^M{}_{Q,}{}^N{}_P+\b_n\d^M_Q\d^N_P=Y^{MN}{}_{PQ}-\d^M_P\d^N_Q$ 
projects on the adjoint of $E_{n(n)}\times\RR$, so that the red factor
    becomes a parameter for an ${\frak e}_n\oplus\RR$ transformation.

The transformations form an
``algebra'' for $n\leq7$: 
$$
[\LL_U,\LL_V]W^M=\LL_{[U,V]}W^M\;,\eqn
$$
where the ``Courant bracket'' is 
$[U,V]^M=\fr2(\LL_UV^M-\LL_VU^M)$, provided that
the derivatives fulfil a ``{\it section constraint}''.

The section constraint ensures that fields locally depend only on
an $n$-dimensional subspace of the coordinates, on which a $GL(n)$
subgroup acts. It reads $Y^{MN}{}_{PQ}\*_M\ldots\*_N=0$, or
$$
(\*\otimes\*)|_{\overline R_2}=0\;.\eqn
$$
For $n\geq8$ more local transformations, so called ``ancillary
transformations'' [\CederwallPalmkvistExtendedGeometry] emerge, which
are constrained local transformations 
in ${\frak g}$.

\vskip2\parskip
\ruledtable
$n$|$R_1$|$R_2$\crthick
$3$|$({\bf3},{\bf2})$|$({\bf\overline3},{\bf1})$\cr
$4$|$\bf10$|$\bf\overline5$\cr
$5$|$\bf16$|$\bf10$\cr
$6$|$\bf27$|$\bf\overline{27}$\cr
$7$|$\bf56$|$\bf133$\cr
$8$|$\bf248$|${\bf1}\oplus{\bf3875}$
\endruledtable
\vskip2\parskip
\centerline{\it Table 2: A list of $R_1$ and $R_2$ for different $E_n$.}
\vskip2\parskip

\Figure\RTwoFigure{\epsffile{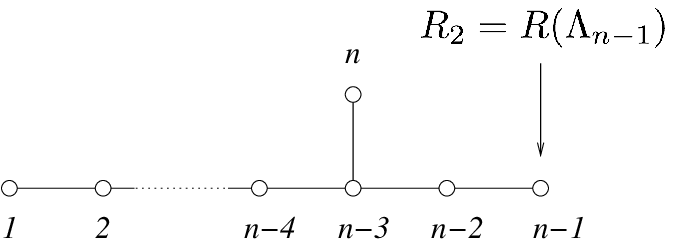}}{The module $R_2$.}


The interpretation of the section condition is that the momenta
locally are chosen so that they may span a linear subspace of cotangent space
with maximal dimension, such that any pair of covectors $p$, $p'$ in
the subspace fulfil $(p\otimes p')|_{\overline R_2}=0$. 

The corresponding statement for double geometry is
$\eta^{MN}\*_M\otimes\*_N=0$, where $\eta$ is the $O(d,d)$-invariant
metric. The maximal linear subspace is a $d$-dimensional isotropic
subspace, and it is determined by a pure spinor $\Lambda$. Once a
$\Lambda$ is chosen, the section condition can be written 
$\Gamma^M\Lambda\*_M=0$.
An analogous linear construction can be performed in the exceptional
setting.
The section condition in double geometry derives from the level
matching condition in string theory.     
Locally, supergravity is recovered. Globally, non-geometric solutions
are also obtained.

There is a universal form
[\PalmkvistBorcherds,\ENinePaper,\CederwallPalmkvistExtendedGeometry]
of the generalised diffeomorphisms
for any Kac--Moody
algebra and choice of coordinate representation.
Let the coordinate representation be $R(\lambda)$, for $\lambda$ a
fundamental weight dual to a simple root $\alpha$ (the construction
can be made more general). Then 
$$
\sigma Y=-\eta_{AB}T^A\otimes T^B+(\lambda,\lambda)+\sigma-1\;,\eqn
$$
where $\eta$ is the Killing metric and $\sigma$ the permutation
operator, $\sigma(a\otimes b)=(b\otimes a)\sigma$.

This follows from the existence of a solution to the section
constraint in the form of a linear space:

\item{$\bullet$}Each momentum must be in the
minimal orbit. Equivalently, $p\otimes p\in\overline{R(2\lambda)}$.

\item{$\bullet$}{Products of different momenta may contain
$\overline{R(2\lambda)}$ and $\overline{R(2\lambda-\alpha)}$, where
$R(2\lambda-\alpha)$ is the highest representation in the
antisymmetric product.}
Expressing these conditions in terms of the quadratic Casimir gives
the form of $Y$.

I will skip the detailed description of the generalised gravity. It
effectively provides the local dynamics of gravity and 3-form, which
are encoded by a vielbein $E_M{}^A$ in the coset
$(E_{n(n)}\times\RR)/K(E_{n(n)})$.

\vskip2\parskip
\ruledtable
$n$|$E_{n(n)}$|$K(E_{n(n)})$\crthick
$3$|$SL(3)\times SL(2)$|$SO(3)\times SO(2)$\cr
$4$|$SL(5)$|$SO(5)$\cr
$5$|$Spin(5,5)$|$(Spin(5)\times Spin(5))/\ZZ_2$\cr
$6$|$E_{6(6)}$|$USp(8)/\ZZ_2$\cr
$7$|$E_{7(7)}$|$SU(8)/\ZZ_2$\cr
$8$|$E_{8(8)}$|$Spin(16)/\ZZ_2$\cr
$9$|$E_{9(9)}$|$K(E_{9(9)})$
\endruledtable
\vskip2\parskip
\centerline{\it Table 3: A list of compact subgroups.}
\vskip2\parskip

The T-duality case is described by a generalised metric 
in the coset
$O(d,d)/(O(d)\times O(d))$, parametrised by the ordinary metric and
$B$-field. 

With some differences from ordinary geometry, one can go through the
construction of connection, torsion, metric compatibility etc.,
and arrive at generalised Einstein's equations encoding the equations
of motion for all fields. (This has been done for $n\leq8$.)

For $n\geq8$, the coset $E_{n(n)}/K(E_{n(n)})$ contains higher mixed
tensors that do not carry independent physical degrees of
freedom. They are removed by ancillary transformations that
arise in the commutator between generalised diffeomorphisms
[\HohmSamtlebenEhlers,\HohmSamtlebenIII,\CederwallRosabal,\ENinePaper,\CederwallPalmkvistExtendedGeometry].

One may introduce (local) supersymmetry. In the case of T-duality, the
superspace is based on the fundamental representation of an
orthosymplectic supergroup $OSp(d,d|2s)$. The exceptional cases are
unexplored, but will be based on $\infty$-dimensional superalgebras
[\CederwallDoubleSuperGeometry].

The generalised diffeomorphisms do not satisfy a Jacobi identity. On
general grounds, it can be shown that the ``Jacobiator''
$$
[[U,V],W]+\hbox{cycl}\neq0\;,\eqn
$$
but is
proportional to 
$([U,V],W)
+\hbox{cycl}$, where
$(U,V)=\fr2(\LL_UV+\LL_VU)$.

It is important to show that the Jacobiator in some sense is
   trivial. It turns out that $\LL_{(U,V)}W=0$ (for $n\leq7$), and the
interpretation is that it is a gauge transformation with a parameter
representing reducibility (for $n\leq6$). (The limits on $n$ in the
statements here are
due to non-covariance of the derivative arising at some point in the
tensor hierarchy, see below. I will not go into details.)

In double geometry, this reducibility is just the scalar reducibility
of a gauge transformation: $\delta B_2=d\lambda_1$, with the
reducibility $\delta\lambda_1=d\lambda'_0$. 

In exceptional geometry, the reducibility turns out to be more
complicated, leading to an infinite (but well defined) reducibility,
containing the modules of tensor hierarchies, and providing a natural
generalisation of forms (having connection-free covariant derivatives).

The reducibility continues, and there are ghosts at all levels $>0$.
The representations are those of a ``tensor hierarchy'', the sequence
of representations $R_n$ of $n$-form gauge fields in the dimensionally
reduces theory.
$$
R_1
\buildrel\partial\over\longleftarrow
R_2
\buildrel\partial\over\longleftarrow
R_3
\buildrel\partial\over\longleftarrow
\ldots\eqn
$$
Example, $n=5$:
$$
{\bf16}
\buildrel\partial\over\longleftarrow
{\bf10}
\buildrel\partial\over\longleftarrow
\overline{\bf16}
\buildrel\partial\over\longleftarrow
{\bf45}
\buildrel\partial\over\longleftarrow
\overline{\bf144}
\buildrel\partial\over\longleftarrow
\ldots\eqn
$$
$$
16-10+16-45+144-\ldots=11\eqn
$$
(suitably regularised),
which is the number of degrees of freedom of a pure spinor.

The representations $\{R_n\}_{n=1}^\infty$ agree with
[\CederwallPalmkvistBorcherds] 
\item{$\bullet$}The ghosts for a ``pure spinor'' constraint (a
constraint implying an object lies in the minimal orbit);
\item{$\bullet$}The positive levels of a Borcherds superalgebra
${\cal B}(E_n)$.

\Figure\BorcherdsFigure{\epsffile{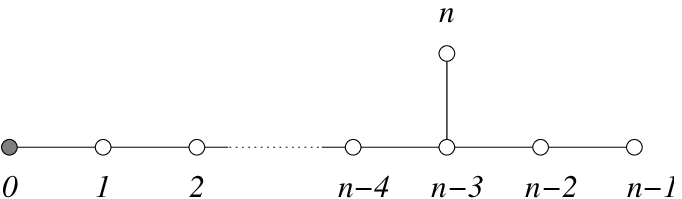}}{Dynkin diagram for ${\cal B}(E_n)$.}

Indeed, the denominator appearing in the
denominator formula for ${\cal B}(E_n)$ is identical to
the partition function of a ``pure spinor''
[\CederwallPalmkvistBorcherds].

${\cal B}(D_n)\approx {\frak osp}(n,n|2)$

${\cal B}(A_n)\approx {\frak sl}(n+1|1)$

$$
\ldots
\buildrel\partial\over\longleftarrow
R_{-1}
\buildrel\partial\over\longleftarrow
R_0
\buildrel\partial\over\longleftarrow
\underbrace{
R_1
\buildrel\partial\over\longleftarrow
R_2
\buildrel\partial\over\longleftarrow
\ldots
\buildrel\partial\over\longleftarrow
R_{8-n}
}_{\hbox{covariant}}
\buildrel\partial\over\longleftarrow
R_{9-n}
\buildrel\partial\over\longleftarrow
R_{10-n}
\buildrel\partial\over\longleftarrow
\ldots\eqn
$$

The modules $R_1,\ldots,R_{8-n}$ behave like forms. The ``exterior
derivative'' is connection-free (for a torsion-free connection), and
there is a wedge product [\CederwallI].

The modules show a symmetry: $R_{9-n}=\overline R_n$.
There is another extension to negative levels that respects this
symmetry, and seems more connected to geometry: tensor hierarchy
algebras [\PalmkvistTensor,\CarboneCederwallPalmkvist]

In the classification of finite-dimensional
superalgebras by Kac, there is a special
class, ``Cartan-type superalgebras''.
The Cartan-type superalgebra $W(n)$, which I prefer to call
$W(A_{n-1})$, is asymmetric between positive and negative levels, and
(therefore) not defined through generators corresponding to simple
roots and Serre relations. 

{\it $W(A_{n-1})$ is the superalgebra of derivations on the superalgebra of
(pointwise) forms in $n$ dimensions.}

Any operation $\omega\rightarrow\Omega\wedge\imath_V\omega$, where
$\Omega$ is a form and $V$ a vector, belongs to $W(A_{n-1})$. A basis
is given by

\vskip2\parskip
\ruledtable
$\hbox{level}=1$|$\imath_a$\cr
$\hfill0$|$e^b\imath_a$\cr
$\hfill-1$|$e^{b_1}e^{b_2}\imath_a$\cr
$\hfill-2$|$e^{b_1}e^{b_2}e^{b_3}\imath_a$\cr
$\hfill\ldots$|$\ldots$
\endruledtable
\vskip2\parskip
\centerline{\it The level decomposition of $W(A_{n-1})$.}
\vskip2\parskip

A subalgebra $S(A_{n-1})$ contains traceless tensors.
The positive levels agree with ${\cal B}(A_{n-1})\approx {\frak sl}(n|1)$.
Note that the representations of torsion and torsion Bianchi identity
appear at levels $-1$ and $-2$.

In spite of the absence of a Cartan involution, there is a way to give
a systematic Chevalley--Serre presentation of the superalgebra, based
on the same Dynkin diagram as the Borcherds superalgebra
[\CarboneCederwallPalmkvist].

\Figure\WgFigure{\epsffile{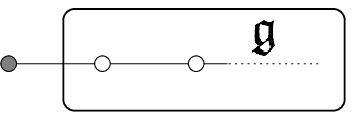}}{Dynkin diagram for ${\cal B}({\frak
g})$ and $W({\frak g})$.}

The construction can be extended to $W(D_n)$, and, most interestingly,
$W(E_n)$ (and the corresponding $S({\frak g})$).
The statements about torsion and Bianchi identities remain true (but
we still lack a good geometric argument).

Back to the Jacobi identity. Expressed in terms of a fermionic ghost
in $R_1$,
$$
[[c,c],c]\neq0\;.\eqn
$$
How is this remedied? The most general formalism for gauge symmetries
is the Batalin--Vilkovisky formalism, where everything is encoded in
the master equation $(S,S)=0$.

If transformations are field-independent, one may consider the ghost
action consistently.
An $L_\infty$ algebra is a (super)algebraic structure which provides a
perturbative solution to the master equation.

Let $C$ denote {\it all} ghosts. Then the master equation states the
nilpotency of a transformation
$$
\delta C=(S,C)=\*C+[C,C]+[C,C,C]+[C,C,C,C]+\ldots\eqn
$$
The identities that need to be fulfilled are:
$$
\eqalign{
&\*^2C=0\komma\cr
&\*[C,C]+2[\*C,C]=0\komma\cr
&\*[C,C,C]+2[[C,C],C]+3[\*C,C,C]\komma\cr
&\ldots\cr
}\eqn
$$
Assuming $\*c=0$, the non-vanishing of $[[c,c],c]$ can be compensated
by the derivative of an element in $R_2$ (representing reducibility).
One needs to introduce a 3-bracket
$$
[c,c,c]\in R_2\;.\eqn
$$
Then, there are more identities to check.

For double field theory, a 3-bracket is enough [\HohmZwiebachLInfinity].

For exceptional field theory, there are signs, that one will in fact
obtain arbitrarily high brackets [\BermanCederwallStrickland].
There are also other issues concerning the non-covariance outside the
``form window''. I will not go into detail.

In conclusion,
the area has rich connections to various areas of pure mathematics,
some of which are under investigation:

\item{$\bullet$}Group theory and representation theory
\item{$\bullet$}Minimal orbits
\item{$\bullet$}Superalgebras
\item{$\bullet$}Cartan-type superalgebras
\item{$\bullet$}Infinite-dimensional (affine, hyperbolic,...) Lie algebras
\item{$\bullet$}Geometry and generalised geometry
\item{$\bullet$}Automorphic forms
\item{$\bullet$}$L_\infty$ algebras
\item{$\bullet$}...

There are many open questions:

\item{$\bullet$}Can the formalism be continued to $n>9$? The
transformations work for \eg\ $E_{10}$
[\CederwallPalmkvistExtendedGeometry], and there seems to be no reason (other
than mathematical difficulties) that it stops there.
Is there a connection to the ``$E_{10}$ proposal'' [\DamourHenneauxNicolaiII]
with emergent space?

\item{$\bullet$}{\it Geometry from algebra?}
What is the precise geometric relation between the tensor
hierarchy algebra and the generalised diffeomorphisms?

\item{$\bullet$}{\it Superspace/supergeometry?} And some formalism
generalising that of 
pure spinor superfields, that manifests supersymmetry?

\item{$\bullet$}{\it The section constraint:} Can it be lifted, or
dynamically generated?

\item{$\bullet$}What can be learnt about the full string theory/M-theory?

\item{$\bullet$}$\ldots$ ?

Thank you for your attention.

\refout







\end